\documentclass[twocolumn]{aastex61}
\usepackage{amsmath}

\begin{document}


\title{Initial Mass Function Variability (or not) Among \\ Low-Velocity Dispersion, Compact Stellar Systems}

\author{Alexa Villaume}
\affiliation{Department of Astronomy \& Astrophysics, University of California Santa Cruz, 1156 High Street, Santa Cruz, CA 95060, USA}
\author{Jean Brodie}
\affiliation{Department of Astronomy \& Astrophysics, University of California Santa Cruz, 1156 High Street, Santa Cruz, CA 95060, USA}
\author{Charlie Conroy}
\affiliation{Harvard-Smithsonian Center for Astrophysics, Cambridge, MA 02138, USA}
\author{Aaron J. Romanowsky}
\affiliation{Department of Physics \& Astronomy, San Jos{\`e} State University, One Washington Square, San Jose, CA 95192, USA}
\affiliation{Department of Astronomy \& Astrophysics, University of California Santa Cruz, 1156 High Street, Santa Cruz, CA 95060, USA}
\author{Pieter van Dokkum}
\affiliation{Yale University Astronomy Department, New Haven, CT 06511, USA}

\begin{abstract}

Analyses of strong gravitational lenses, galaxy-scale kinematics, and absorption line stellar population synthesis (SPS) have all concluded that the stellar initial mass function (IMF) varies within the massive early-type galaxy (ETG) population. However, the physical mechanism that drives variation in the IMF is an outstanding question. Here we use new SPS models to consider a diverse set of compact, low-velocity dispersion stellar systems: globular clusters (GCs), an ultra-compact dwarf (UCD), and the compact elliptical (cE) galaxy M32. We compare our results to massive ETGs and available dynamical measurements. We find that the GCs have stellar mass-to-light ratios (M/L) that are either consistent with a Kroupa IMF or are slightly bottom-light while the UCD and cE have mildly elevated M/L. The separation in derived IMFs for systems with similar metallicities and abundance patterns indicates that our SPS models can distinguish abundance and IMF effects. Variation among the sample in this paper is only $\sim 50\%$ in normalized M/L compared to the $\sim 4\times$ among the ETG sample. This suggests that metallicity is not the sole driver of IMF variability and additional parameters need to be considered.
\end{abstract}

\section{Introduction}

The assumption of a universal stellar initial-mass function (IMF) has been a cornerstone of stellar population and galaxy evolution studies for decades. Nevertheless, there has been much observational effort to test and challenge this assumption. The work done in nearby systems where it is possible to measure resolved star counts is extensive \citep[see Ch. 9 in][and references therein]{kroupa2013}. Since the discovery of surface gravity sensitive absorption features  \citep[e.g.,][]{wing1969} the measurement of the IMF in systems beyond the reach of resolved star counts has been possible. In principle, these lines can measure the ratio of giant-to-dwarf stars in integrated light, which can be used as an IMF proxy \citep[e.g.,][]{cohen1978, faber1980, kroupa1994}.

In practice, only in recent years have the stellar population synthesis (SPS) model precision and near-infrared (near-IR) data quality reached the point where it is to possible measure the dwarf-to-giant ratio. \citet{cenarro2003} found that age and metallicity effects alone could not explain the variations in CaT strength in a sample of early-type galaxies (ETGs) and tentatively attributed it to IMF variability. More recent work \citep[e.g.,][]{vd2010, spin2011, cvd2012, fer2013, mn2015} has made progress on making quantitative statements about the relative number of giant and dwarfs stars. The results from SPS modeling broadly agree with investigations using gravitational lensing and kinematics \citep[e.g.,][]{treu2010, cappellari2013}. However, there remain inconsistencies from the different methods on an object-by-object basis \citep[][]{smith2014}.

There is not yet a clear physical mechanism driving IMF variability. Metallicity has become a possibility from recent observational work \citep[][]{mn2015,vd2016} but velocity dispersion ($\sigma$) and $\alpha$-element abundances also correlate with IMF variation \citep[][]{cvd2012, labarbera2013}. Furthermore, there are still unexplained complications in the emerging picture of IMF variability. \citet{newman2016} demonstrated that even high-velocity dispersion ETGs can have MW IMFs, and, furthermore, it is not yet clear how IMF variability conforms to the expectations from chemical evolution and star-formation measurements \citep[e.g.,][]{mn2016}.
  
Most integrated light probes of the IMF focused on ETGs and so have only looked at IMF variations in relatively narrow regions of parameter space. To better constrain IMF variations as a function of the physical characteristics of the stellar population we need to push IMF studies to the extremes of parameter space. Ultracompact dwarfs (UCDs) are extremely dense objects that can have high dynamical mass-to-light ratio values (M/L)$_{\rm dyn}$ \citep[e.g.,][]{mieske2013}. Globular clusters (GCs) are conventionally thought to have \citet{kroupa2001} (MW) IMF. However, \citet{strader2011} found a trend of decreasing (M/L)$_{\rm dyn}$  of M31 GCs as a function of metallicity, in disagreement with the expectation from a MW IMF.

Whether UCDs and GCs actually have variable IMFs and, if so, what the shape is, is still being debated \citep[][]{jera2017}. \citet{dabringhausen2012} took an overabundance of X-ray binaries in a sample of Fornax UCDs as evidence that those UCDs produced more massive stars than expected from a Kroupa IMF. \citet{marks2012} used the gas-expulsion timescale of a sample of UCDs and GCs  to predict that the IMF would create more massive stars with increasing density. However, \citet{pandya2016} analyzed 336 spectroscopically confirmed UCDs across 13 host systems and found an X-ray detection fraction of only $\sim3\%$. \citet{zonoozi2016} showed that the combination of a variable IMF and removal of stellar remnants could plausibly explain the (M/L)$_{\rm dyn}$ trend in the M31 GCs.

 Fitting the integrated light of UCDs and GCs with SPS models is needed to obtain a more direct measurement of the IMF shape. One caveat is that GCs can be strongly influenced by dynamical evolution, i.e., mass-segregation and evaporation of low-mass stars. For the low-mass stars the ``initial'' mass function is not being measured, but rather the ``present-day'' mass function (PDF). However, this should not be a concern for high mass GCs or UCDs, the PDF is expected to closely resemble the IMF owing to long relaxation times \citep[see eq. 17 in][]{portegies2010}. 

In this paper we present a pilot study of stellar mass-to-light ratios, (M/L)$_{*}$, of various compact stellar systems (CSSs): M59-UCD3 \citep{sandoval2015}, three M31 GCs that span a large range of metallicity, and the compact elliptical (cE) M32. For the first time we fit the spectra of the individual objects with flexible SPS models that allow IMF variability. 

\section{Observations and Data}

 \begin{figure}[t]
\includegraphics[width=0.5\textwidth]{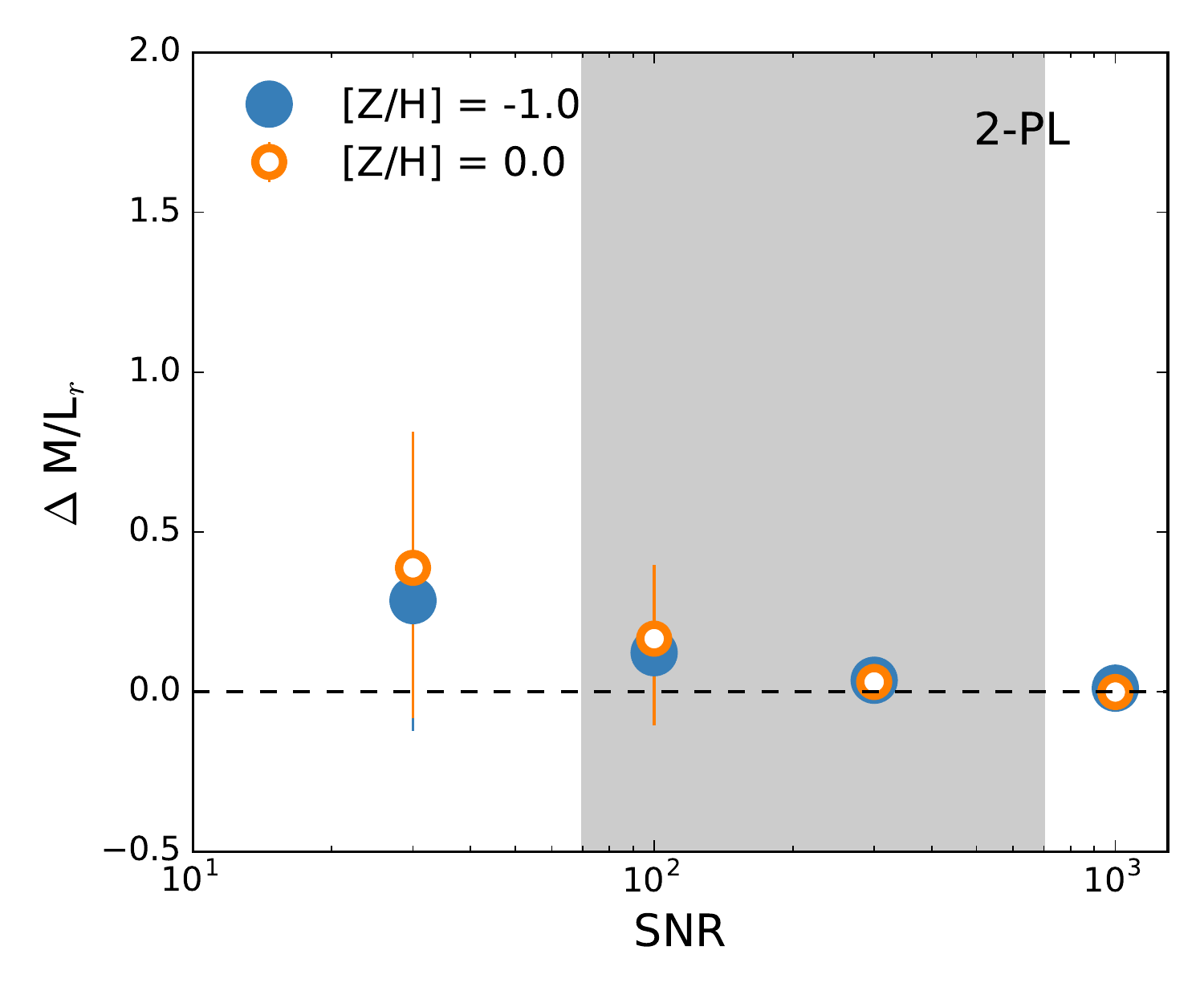}
\caption{Recovery of (M/L)$_{*}$ from mock data as a function of S/N for [$Z$/H] = 0.0 (orange) and [$Z$/H] = $-1.0$ (blue) models. The circles show the median difference between the input (M/L)$_{*}$ and the inferred (M/L)$_{*}$ derived from the fits of 10 realization of mock data. A S/N of $\gtrapprox 100$ is needed to recover the M/L. The grey band shows the range of S/N values in the data.}
\label{figure:snr_ml}
\end{figure}

\begin{figure*}
\includegraphics[width=1.0\textwidth]{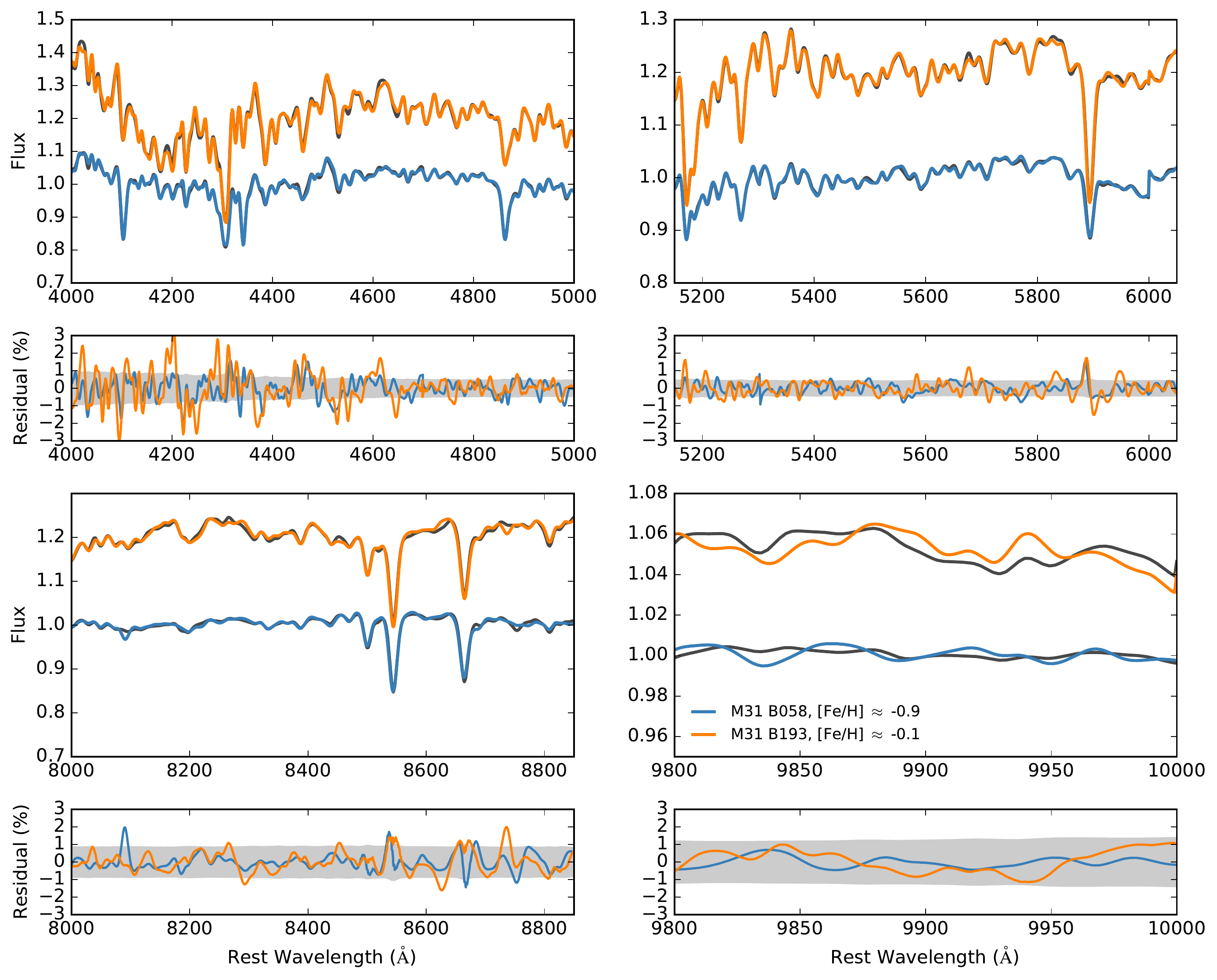}
\caption{(Upper panels) Comparison of best-fit models (grey) and data in key wavelength regions for M31-B193 (metal-rich GC, orange) and M31-B058 (metal-poor GC, blue). (Lower panels) Comparison of the percentage difference between the best-fit model and data regions for M31-B193 and M31-B058. The data have been smoothed and so the pixels are highly correlated. In the grey band we show the uncertainty for one of the GCs, M31-B058, as the uncertainties are comparable. The residuals between metal-rich and metal-poor GC are also comparable.}
\label{figure:gc_residual}
\end{figure*}

All of the objects presented in this paper were observed with LRIS \citep{oke1995}, a dual-arm spectrograph, on the Keck I telescope on Maunakea, Hawaii.

The data for one metal-poor (MP) GC (M31-B058), two metal-rich (MR) GCs (M31-B163 and M31-B193), and M59-UCD3 were obtained on December 19--20 2014, using the instrument setup and using the same ``special'' long slit discussed in \citet{vd2016} ($0.7^{\prime \prime} \times 290^{\prime \prime}$). Since the objects in this paper are bright and compact we obtained 4 300s exposures using an ABAB pattern where we dithered up and down the slit by 20$^{\prime \prime}$.  

Three exposures of 180 s were taken for M32 on January 2012. The 600 l mm$^{-1}$ grating was used on the blue arm but the same grism as the other objects was used on the red arm. We extracted a spectrum using a square aperture of 0.8$^{\prime \prime}$x0.8$^{\prime \prime}$ ($\approx 3$ pc).

The intrinsic resolution of the the objects in this sample is higher than the models (which are smoothed to a common resolution of $\sigma = 100$ km~s$^{-1}$) so we broadened the spectra in our sample. To have roughly the same dispersion in the red for all objects we broadened the M32 and UCD spectra by 150 km~s$^{-1}$ and the GCs by 200 km~s$^{-1}$.

\section{Modeling}

\subsection{Model Overview}

The methodology we use for fitting the models to data and the parameters fitted are described in detail in \citet{conroy2017b}. The models described in \citet{conroy2017b} (``C2V'' models) are the updated versions of the stellar population models from \citet{cvd2012} (``CvD'' models). The most important update for this paper is the increased metallicity range provided by the Extended IRTF library \citep{villaume2017} and metallicity-dependent response functions.

We explore the parameter space using a Fortran implementation of emcee \citep{fm2013}, which uses the affine-invariant ensemble sampler algorithm \citep{goodman2010}. We use 512 walkers, 25,000 burn-in steps, and a production run of 1,000 steps for the final posterior distributions. 

We perform full-spectrum fitting. We continuum normalize the models by multiplying them by higher-order polynomials to match the continuum shape of the data. 

We sample the posteriors of the following parameters: redshift and velocity dispersion, overall metallicity,  a two component star formation history (two bursts with free ages and relative mass contribution), 18 individual elements, the strengths of five emission line groups,  fraction of light at 1$\mu m$ contributed by a hot star component, two higher order terms of the line-of-sight velocity distribution, and nuisance parameters for the data (normalization of the atmospheric transmission function, error and sky inflating terms).\footnote{Models fitted with only a single age and excluding the emission lines made a negligible effect on the inferred parameters for the GCs.}

Additionally, we fit for the slopes of a two component power-law (break point at $0.5$ M$_\odot$):

\[
 \xi(m) = {\text d}N/{\text d}m_* =
  \begin{cases} 
   k_1m^{-\alpha_1} \text{ for } 0.08 < m < 0.5,\\
   k_2m^{-\alpha_2} \text{ for } 0.5 < m < 1.0\text{, and }\\
   k_3m^{-2.3} \text{ for } \geq 1.0.
  \end{cases}
\]

\noindent For a MW IMF $\alpha_1 = 1.3$ and $\alpha_2 = 2.3$. The IMF above 1.0M$_\odot$ is assumed to have a  \citet{salpeter1955} slope. The $k_i$'s are normalization constants that ensure continuity of the IMF. The upper mass limit is 100M$_\odot$ and the low-mass cutoff, $m_c$, is fixed at 0.08M$_\odot$.  In this paper we present our IMF results in terms of (M/L)$_*$. The mass of the stellar population is calculated from the best inferred slopes of the IMF and stellar remnants are included in the final mass calculation following \citet{conroy2009}. A stellar population is considered bottom-heavy, an overabundance of low-mass stars, if the exponents on the first two terms are larger than the MW IMF and is considered bottom-light, a paucity of low-mass stars, if they are less than those values.

\subsection{Mock Data Demonstrations}
 \begin{deluxetable*}{llcccccc}
\tablecaption{List of objects and associated physical parameters. \label{table:obj_info}}
\tablehead{
 \colhead{Object}  & \colhead{S/N} & \colhead{$\sigma$} & \colhead{[Fe/H]} & \colhead{Age} & \colhead{[Mg/Fe]} & \colhead{M/L$_V$} & \colhead{ M/L$_V$} \\
\colhead{}  & \colhead{${\rm \AA}^{-1}$} &  \colhead{(km~s$^{-1}$)} & \colhead{} & \colhead{(Gyr)} & \colhead{} & \colhead{2 PL} & \colhead{MW} 
}
\startdata
M32 & 730\tablenotemark{*} & 75\tablenotemark{a} & 0.15$_{-0.01}^{+0.01}$ &       2.98$_{-0.06}^{+0.05}$ &       0.02$_{-0.01}^{+0.04}$ &       2.4$_{-0.64}^{+0.64}$ &       1.63$_{-0.03}^{+0.03}$ \\ 
M59-UCD3 & 70 &  70\tablenotemark{b} &  0.01$_{-0.01}^{+0.01}$ &       7.7$_{-0.48}^{+0.49}$ &       0.18$_{-0.01}^{+0.01}$ &       5.1$_{-1.17}^{+0.87}$ &       2.98$_{-0.1}^{+0.11}$ \\
M31-B163 &  100 &  21\tablenotemark{c} & $-0.18_{-0.01}^{+0.01}$ &       11.37$_{-0.61}^{+0.7}$ &       0.21$_{-0.01}^{+0.01}$ &       3.61$_{-0.49}^{+0.59}$ &       3.34$_{-0.11}^{+0.12}$ \\ 
M31-B193 & 250 &  19\tablenotemark{c} & $-0.11_{-0.01}^{+0.01}$ &       9.7$_{-0.45}^{+0.54}$ &       0.24$_{-0.01}^{+0.01}$ &       2.69$_{-0.2}^{+0.43}$ &       3.16$_{-0.1}^{+0.09}$ \\
M31-B058 & 120 & 23\tablenotemark{c} & $-0.96_{-0.01}^{+0.01}$ &       6.92$_{-0.1}^{+0.09}$ &       0.37$_{-0.02}^{+0.02}$ &       1.38$_{-0.08}^{+0.07}$ &       1.54$_{-0.01}^{+0.01}$ \\ 
\enddata
\tablecomments{Mean best inferred value for each parameter is shown with 1$\sigma$ statistical uncertainty. Values were determined with our models and fitting procedure, as described in Section 3.1}. The second to last column are the (M/L)$_{*}$ values where the IMF was allowed to vary as a two component power-law IMF and the last column is the  (M/L)$_{*}$ values where the IMF was fixed to a Kroupa IMF. 
\tablenotetext{*} {Although the S/N is high it was cloudy at the time of observation so there is additional uncertainty in the data not represented by Poisson statistics.}
\tablenotetext{a} {\citet{gult2009}}
\tablenotetext{b} {\citet{janz2016}}
\tablenotetext{c} {\citet{strader2011}}
\end{deluxetable*}

 To test our ability to recover (M/L)$_{*}$ from the data, we synthesize mock spectra by assuming a Salpeter IMF, adding different amounts of noise, and then use our models and fitting procedures to derive $\Delta$ M/L$_{*}$. We show $\Delta$ M/L$_{*}$ for mock spectra with solar, [$Z$/H] $= 0.0$ (orange), and sub-solar, [$Z$/H] $= -1.0$ (blue). For each S/N and metallicity value we create 10 mock spectra with fixed S/N per ${\rm \AA}$ over the wavelength range $0.4-1.015$$\mu m$, a velocity dispersion of $250$ km~s$^{-1}$, and an age of 10 Gyr. The abundance patterns of the mock spectra are solar scaled \citep[e.g.,][]{choi2016} and the nuisance parameters are set to zero. The points shown in Figure~\ref{figure:snr_ml} are the median values of the differences between the input (M/L)$_{*}$ and the derived (M/L)$_{*}$ from the inferred IMF parameters for each metallicity and S/N pair. The uncertainties shown are the median statistical uncertainties of the recovered values.

For solar metallicity the models recover (M/L)$_{*}$ when the S/N $\gtrapprox 100{\rm \AA^{-1}}$. A similar trend is also seen in the low-metallicity mock data. While not a significant difference, it is somewhat counterintuitive that the (M/L)$_{*}$ at the low-S/N regime is better recovered for the low-metallicity mocks. It could be that in the low-S/N regime weaker metal lines help distinguish IMF effects. Below S/N $\sim 100{\rm \AA^{-1}}$  there will be large uncertainty and bias in the (M/L)$_{*}$ measurement. The bias exists in the low-S/N regime because the priors become important and the truth is at the edge of the prior. The measurements are less sensitive to S/N if the true $m_c$ is higher \citep[see][for details]{conroy2017a}. As discussed in \citet{conroy2017a} the S/N requirements for allowing $m_c$ to vary is even higher than what is shown in Figure~\ref{figure:snr_ml}. Most of the data in this paper do not meet the S/N requirements for this type of parametrization.

\section{Results}

\subsection{Basic Stellar Population Characteristics}

In the upper panels of Figure~\ref{figure:gc_residual} we compare the best-fit models (grey) and data for M31-B193 (orange), a metal-rich (MR) GC, and M31-B058 (blue), a metal-poor (MP) GC. In the lower panels we show the percentage difference between the models and data. The uncertainty for M31-B058 is shown by the grey band (the uncertainty for M31-B193 is comparable). The CvD models would not have been able to fit M31-B058 because of the limited metallicity range, but with the C2V models the residuals between MP and MR GC are comparable and small. 

In Table~\ref{table:obj_info} we show the best inferred median values for [Fe/H], mass-weighted age, [Mg/Fe], and the (M/L)$_{*}$ in Johnson $V$ where we have and have not allowed the IMF to vary from Kroupa. Our stellar parameters are broadly consistent with previous work on these objects. From deep {\it HST}/ACS imaging of M32 \citet{monachesi2012} inferred two dominant populations, one 2--5 Gyr and metal-rich and an older population, $\sim 7$ Gyr. Our inferred age skews young as the integrated light observations are almost certainly dominated by the young population. \citet{monachesi2012} determined near-solar mass- and light-weighted metallicities for M32. Our inferred metallicity is slightly more metal-rich than that. \citet{janz2016} used Lick indices on M59-UCD3 and found [$Z$/H] $= 0.15 \pm 0.10 $. Converting our value for [Fe/H] to [$Z$/H] \citep{trager2000} we get  [$Z$/H] $\approx 0.2$, consistent with the  \citet{janz2016} value. Furthermore, our inferred values for M59-UCD3 are consistent with those presented in \citet{sandoval2015} with a spectrum from a different instrument and an earlier iteration of our models.

Our inferred ages for M31-B163 and B193 are consistent with the ages derved by \citet{colucci2014}. This is particularly striking since \citet{colucci2014} worked with high-resolution data and a completely different analysis technique. The age for M31-B058 is young for a GC but is consistent with previous work in modeling integrated light of MP GCs \citep[][]{graves2008}. In the case of M31-B058 there is a moderate blue horizontal branch that could be boosting the strength of the Balmer lines \citep{rich2005}. 

\subsection{The IMF}

\begin{figure*}
\includegraphics[width=1.0\textwidth]{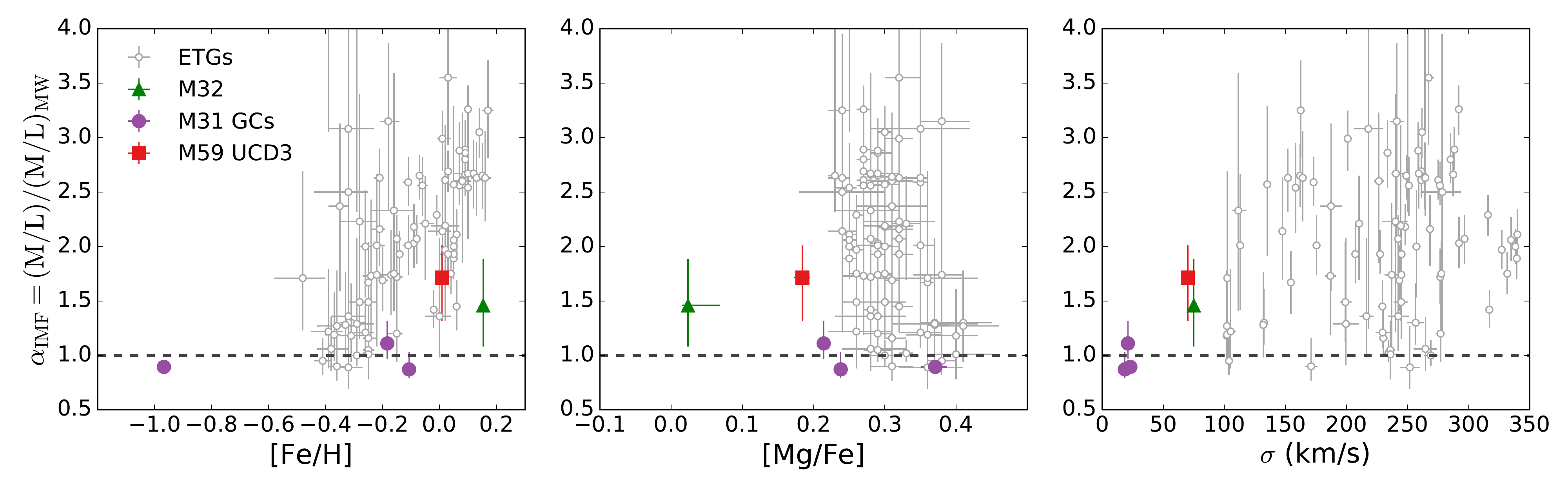}
\caption{The IMF mismatch parameter plotted against [Fe/H] (left), [Mg/Fe], (middle), and  $\sigma$ (right) for the two component power-law IMF. Values shown are for M59-UCD3 (red squares), the M31 GCs (purple cricles), and M32 (green triangles). We show the full sample of ETG local values from \citet{vd2016} (open grey).}
\label{figure:alpha_imf1}
\end{figure*}

For our main analysis we define the ``IMF mismatch'' parameter, $\alpha_{\rm IMF}$. This parameter is the ratio of (M/L)$_{*}$ where we have fitted for the IMF,  to (M/L)$_{*}$ where we have assumed a MW IMF. In Figure~\ref{figure:alpha_imf1} we show $\alpha_{\rm IMF}$ plotted against [Fe/H] (left),  [Mg/Fe] (middle), and velocity dispersion ($\sigma$, right) for all the objects in our sample: the M31 GCs (purple), M59-UCD3 (red),  M32 (green). We supplement our data set with the ETG data from \citet{vd2016} (grey, open circle) with the same instrumental and model setups. 
 
\begin{figure*}
\includegraphics[width=1.0\textwidth]{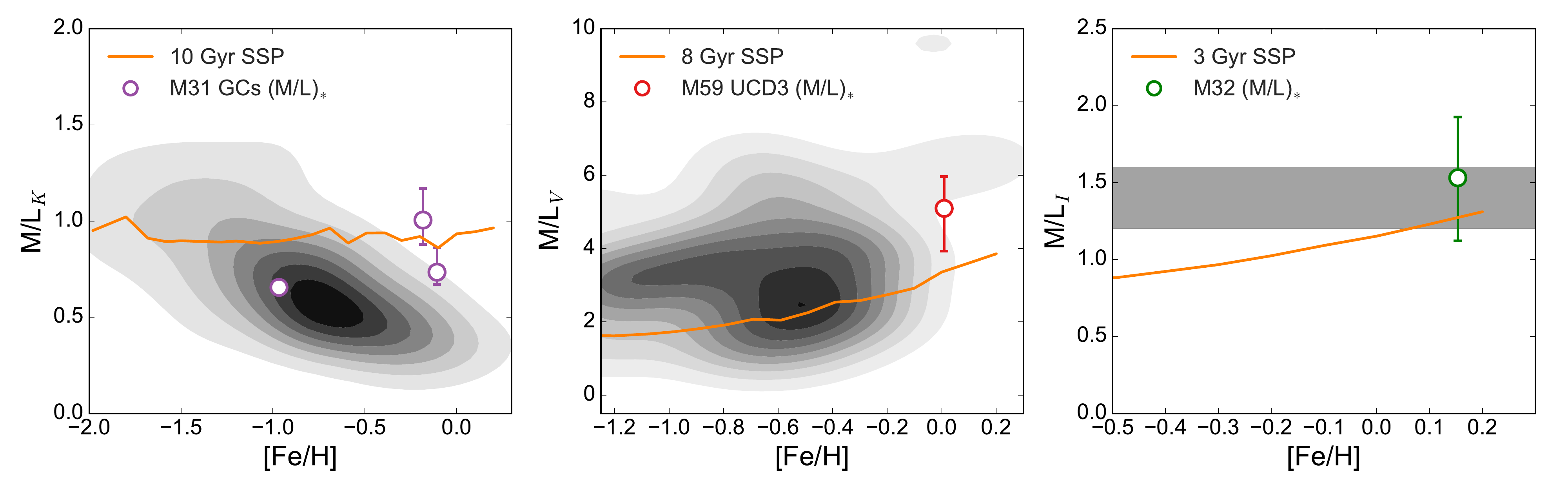}
\caption{Comparison of (M/L)$_{\rm dyn}$ (grey) to (M/L)$_{\rm *}$ values for M31 GCs (left, purple), M59-UCD3 (middle, red), and M32 (right, green). In each panel we show the metallicity-dependent (M/L)$_{*}$ predicted from SSPs with Kroupa IMF and solar-scaled abundance patterns. The ages of the SSPs (orange line) were chosen to approximate the inferred ages from our full-spectrum fitting. Our inferred (M/L)$_{\rm *}$ values for M59-UCD3 and M32 are consistent with available (M/L)$_{\rm dyn}$ measurements. There remain inconsistencies between the dynamical and stellar measurements at high metallicity for the M31 GCs.}
\label{figure:dyn_mass_compare}
\end{figure*}

 In Figure~\ref{figure:dyn_mass_compare} we compare our (M/L)$_{*}$ measurements with available (M/L)$_{\rm dyn}$ measurements. In the left panel, we show the kernel density estimate (KDE) for [Fe/H] vs. (M/L)$_{\rm dyn}$ for M31 GCs from \citet{strader2011} (contours, darker color indicates higher concentration of objects) along with our (M/L)$_{*}$ for three GCs. Published (M/L)$_{\rm dyn}$ measurements do not currently exist for M59-UCD3. However, in the middle panel we show the KDE of [Fe/H] vs. (M/L)$_{\rm dyn}$ of the sample of UCDs from \citet{mieske2013} (we removed objects that belong to NGC 5128 owing to suspicions of spurious $\sigma$ measurements) and (M/L)$_{*}$ for M59-UCD3. In the right panel of Figure~\ref{figure:dyn_mass_compare} we compare (M/L)$_{\rm *}$ for M32 with (M/L)$_{\rm dyn}$ from \citet{vdB2010} where the grey band represents the lower and upper limits given by the uncertainty. In each panel we show metallicity-dependent (M/L)$_{*}$ predictions using SSPs with MW IMFs and solar-scaled abundance patterns. The ages of the SSPs were chosen to approximate the inferred ages from full-spectrum fitting.

We note the slight discrepancy in Figures~\ref{figure:alpha_imf1}  and \ref{figure:dyn_mass_compare} in how much M32 appears to deviate from a MW IMF. This is due to the fact that the MW IMF in Figure~\ref{figure:alpha_imf1} also accounts for non-solar abundance patterns while the SSPs used to generate the orange lines in Figure~\ref{figure:dyn_mass_compare} do not.

\section{Discussion}

\citet{mcconnell2016} and \citet{ziel2017} computed line indices for a variety of ETGs and claimed that observed line strengths can be explained by abundance variations alone. These studies have driven debates about the extent IMF measurements are affected by the underlying abundance patterns. The M31 GCs are an excellent test bench for the models in this respect since they have similar metallicities and element enhancements as massive ETGs. If the models did conflate metallicity and abundance effects with IMF effects we would expect to find similar (M/L)$_{*}$ enhancements in the M31 GCs. Recovering $\alpha_{\rm IMF} \sim 1$ for the M31 GCs over a wide metallicity range is a strong validation that our models can distinguish IMF and abundance effects. 

Our modeling of the M31 GCs improves upon earlier work in several important ways. \citet{zonoozi2016} did not fit models to data and assumed a top-heavy IMF. \citet{cvd2012} used a stacked spectrum of MR GCs to test the CvD models while making measurements for individual clusters and include a MP GC. The lack of expected dark matter in GCs means that dynamical measurements provide tight constraints on our expectations for (M/L)$_{*}$. This makes the continued discrepancy between dynamical and stellar measurements on the MR end of the M31 GCs troubling.

For the current models $m_c$ is fixed at $0.08M_\odot$ but a higher $m_c$ would lower the inferred (M/L)$_*$ values. \citet{chabrier2014} explored the different theoretical conditions which would create a higher $m_c$, while there is empirical evidence that the IMF in GCs becomes flatter for $< 0.5M_\odot$ \citep[][]{marks2012}, which would mimic an increase in $m_c$. It is not out of the realm of possibility that $m_c$ could differ from our fiducial value. However, it takes increasing $m_c$ to  $0.5 {\rm M}_{\odot}$, an extreme value, to decrease (M/L)$_{\rm *}$ by 35\%, i.e., closer to the locus of the MR (M/L)$_{\rm dyn}$ values. It is premature to make any definitive conclusions but these preliminary results suggest that a variable IMF cannot explain the [Fe/H] vs (M/L)$_{\rm dyn}$ trend for the M31 GCs. \citet{zonoozi2016} were able to achieve better agreement by making {\it ad hoc} adjustments to the retention rates of stellar remnants in the GCs.  Follow-up work with a larger sample and more detailed physical models is required.

The mild bottom-heaviness of M59-UCD3 contrasts with the expectations of \citet{dabringhausen2012} and \citet{marks2012}. That is not to say that our results are in direct contradiction with either study. First, those studies are tracing the  stars and we are tracing the low-mass stars. Second, It is becoming increasingly clear that UCDs as a class encompass a diverse set of objects \citep{janz2016}. Until we have a better understanding of a more comprehensive sample of objects it is premature to make any firm conclusions about how UCDs {\it as a whole} behave.

For the sample presented in this work, the main feature of Figure~\ref{figure:alpha_imf1} is that the CSSs are distinct from the main ETG sample. Though they span large [Fe/H] and [Mg/Fe] ranges, they vary much less in $\alpha_{\rm IMF}$ than the ETG sample. Both M59-UCD3 and M32 have elevated $\alpha_{\rm IMF}$ values but are not on the main [Fe/H]--$\alpha_{\rm IMF}$ trend for massive ETGs. M59-UCD3 is in a cluster of ETG points that also deviate from the main trend. Those points originate from the central regions of just two of the galaxies in the ETG sample: NGC 1600 and NGC 2695. 

The main conclusion of this work is that metallicity is not the sole driver of IMF variability \citep[see][]{mn2015, vd2016}. The right panel of Figure~\ref{figure:alpha_imf1} suggests that velocity dispersion is also associated with IMF variation. This is an important result because different theoretical frameworks will be controlled by different fundamental variables depending on the kind of physics they evoke to fragment gas clouds \citep[see][]{krumholz2014}. By expanding IMF probes into the parameter space that CSSs occupy we can elucidate what these variables are.

 Moreover, it is unclear how theoretical frameworks of star-formation should treat  monolithically formed populations (GCs, some UCDs) as compared with populations that build up over time (some CSSs and ETGs) \citep[see Ch. 13 in][]{kroupa2013}. By measuring the IMFs of CSSs with the same modeling framework that we do for ETGs, we can obtain a self-consistent observational picture of how the IMF manifests in the different types of population. Currently, with our small sample, it is unclear whether the GCs have IMFs that are distinct from the UCDs and cEs (the left and middle panels of Figure~\ref{figure:alpha_imf1}) or are a part of the same continuum (right panel of Figure~\ref{figure:alpha_imf1}).

\acknowledgements

The authors wish to recognize and acknowledge the very significant cultural role and reverence that the summit of Maunakea has always had within the indigenous Hawaiian community. AV is supported in part by a NSF Graduate Research Fellowship. JB acknowledges support from NSF grants AST-1518294 and AST-1616598. CC acknowledges support from NASA grant NNX15AK14G, NSF grant AST-1313280, and the Packard Foundation. AJR was supported by NSF grants AST-1515084 and PHY11-25915, and as a Research Corporation for Science Advancement Cottrell Scholar. The authors would like to thank I. Mart{\'i}n-Navarro, J. Strader, A. Seth, and C. Ahn for helpful discussion and the anonymous referee, who's thoughtful comments improved the quality of this manuscript.

 \software{\texttt{astropy} \citep{2013A&A...558A..33A},\, \texttt{matplotlib} \citep{Hunter:2007},\, \texttt{NumPy}\,\citep{2011arXiv1102.1523V},\, \texttt{Seaborn}\,\citep{SeaBorn}}

\end{document}